# Dynamic Resource Coordination and Interference Management for Femtocell Networks

Yan Liang, Chengling Jiang, Chunliang Yang

**Abstract**

Femtocell is emerging as a key technology to secure the coverage and capacity in indoor environments. However the deployment of a new femtocell layer may originate undesired interference to the whole system. This paper investigates spectrum resource coordination and interference management for the femtocell networks. A resource coordination scheme based on broadcasting resource coordination request messages by the femto mobile is proposed to reduce the system interference.

## 1. Introduction

Femtocells are low-power base stations installed by the consumer to enhance indoor coverage. They have the potential to provide high quality network access to indoor users at low cost, while simultaneously reducing the burden on the whole system. On the air interface, femtocells provide radio coverage of a given cellular standard (e.g., GSM, UMTS, WiMAX, LTE), while the backhaul connection makes use of a broadband connection such as optical fiber or digital subscriber line (DSL). Compared to other techniques for increasing system capacity, such as distributed antenna systems and microcells, the key advantage of femtocells is that there is very little upfront cost to the service provider.

However, since the existing macrocell network is overlaid on femtocell networks utilizing the same set of frequency channels, femto basestation (fBS) can originate severe interference to the macrocell network or other fBSs unless the femtocell network is carefully configured. An approach that completely eliminates cross-layer interference is to divide the licensed spectrum into two parts (orthogonal channel assignment). This way, a fraction of the subchannels would be used by the macrocell layer while another fraction would be used by the femtocells. Although optimal from a cross-layer interference standpoint, this approach is inefficient in terms of spectrum reuse. Therefore, co-channel assignment of the macrocell and femtocell layers seems more efficient and profitable for operators, although far more intricate from the technical point of view.

Since the number and position of the femtocells are initially unknown due to the individualistic nature of the femtocells, a distributed approach to mitigate cross- and co-layer interference, where each cell manages its own subchannels, is more suitable than centralized approach. The femtocells could directly exchange information about their subchannel usage, spectrum needs, and so on. This way, femtocells can be aware of the present actions and future intentions of their neighboring cells, and act accordingly. These messages can be exchanged through backhaul, however there would be latency issues.

In this paper, we propose a resource coordination scheme for the femtocell networks based on broadcasting messages by femto mobile stations (fMS) to reduce system interferences and achieve higher capacities. The rest of this paper is organizes as follows. The resource coordination algorithm is represented in Section 2. Section 3 concludes this paper.

## 2. Resource Coordination Schemes

In this section, a resource coordination scheme is proposed to reduce the system interferences. The main idea is that the fMSs broadcast resource coordination request messages (RRM) to BSs through air interface and fBSs do power control according to the information from RRMs they receive. This resource coordination scheme is a quick approach and could be used before the long term coordination achieved by message exchange through backhaul.

### 2.1 Procedure of Resource Coordination

The detailed procedure of resource coordination is as follows. After an fMS (fMS$_i$) sends out request for data transmission, its serving fBS (fBS$_i$) allocates frequency-time resource to it. Then fMS$_i$ will broadcast its RRM to all the BSs for resource coordination. This RRM includes the following information: target SINR, required resource id, hashed serving base station id and data priority. The fBS which receives the RRM will do power control to limit the power on the required frequency source in order to reduce co-channel interference. After the power control step, each fBS will send out pilot with the controlled power on the required resource. Each fMS which sends RRM receives the pilots from fBSs, calculates its SINR and sends this CQI information to its serving fBS. Then the serving fBS does scheduling and begins data transmission.

### 2.2 Power Control Based on Data Priority

A simple power control approach is designed based on data priority comparison. Each fBS receives $N$ RRMs from different fMSs and compares the data priority information. The fMS with the highest data priority will be assigned full power strength for data transmission; while for other fMSs, power will be shut down to avoid co-channel interference. If there are totally $M$ fMSs ($M>1$) share the same highest data priority, each fMS will have its data transmitted randomly with the chance of $\frac{1}{M}$.

### 2.3 Power Control Based on System Throughput

Instead of turning off the transmission when higher priority RRM is received, we can also optimize the system throughput as:

$$P_c^*(\omega) = \arg\max_{0 \le P_c < P} \left\{ \sum_{\substack{i \in \Gamma_\omega^c \\ i \ne c}} \eta_i \log_2\left(1 + \frac{Ph_i}{P_c h_{ic} + \sigma_i^2}\right) + \eta_c \log_2\left(1 + \frac{P_c h_c}{\sum_{\substack{i \in \Gamma_\omega^c \\ i \ne c}} Ph_{ci} + \sigma_c^2}\right) \right\} \quad (1)$$

where $P$ is the maximum transmit power for fMS, $P_c$ is the controlled data power for fMS$_c$, $\sigma_i^2$ is the receiver noise variance at fMS$_i$, h$_i$ is the channel gain from fBS$_i$ (serving sector of MS$_i$) to fMS$_i$, h$_{ic}$ is the channel gain from fMS$_i$ to fBS$_c$, which is the same as the channel gain from fBS$_c$ to fMS$_i$ because of the reciprocity characteristic of the channel, $\eta_i$ is the data priority of fMS$_i$, and $\Gamma_\omega^c$ is the set of fBSs that are requesting coordination for resource $\omega$ that are received by

fMS$_c$.

The optimization objective in (1) could be re-written as

$$\arg\max_{0\leq P_c<P}\left\{\sum_{\substack{i\in\Gamma_\omega^c\\i\neq c}}\eta_i\log_2\left(1+\frac{1}{\frac{P_ch_{ic}}{Ph_i}+\frac{\sigma_i^2}{Ph_i}}\right)+\eta_c\log_2\left(1+\frac{P_ch_c}{\sum_{\substack{i\in\Gamma_\omega^c\\i\neq c}}Ph_{ci}+\sigma_c^2}\right)\right\} \quad (2)$$

Approximately,

$$\arg\max_{0\leq P_c<P}\left\{\sum_{\substack{i\in\Gamma_\omega^c\\i\neq c}}\eta_i\log_2\left(1+\frac{1}{\frac{P_ch_{ic}}{Ph_i}+\frac{1}{SINR_i}}\right)+\eta_c\log_2\left(1+\frac{P_ch_c}{\sigma_c^2}\right)\right\} \quad (3)$$

where $SINR_i$ is the target SINR of fMS$_i$,

The transmit power for RRM is designed to be $p_{tran}=\min\left\{\frac{P^{RRM}}{h_i},P\right\}$, where $P^{RRM}<P$ is the nominal RRM transmit power, then the received RRM power at fBS$_c$ is $p_{rec,i}=h_{ic}p_{tran}=\min\left\{\frac{P^{RRM}h_{ic}}{h_i},h_{ic}P\right\}$.

Thus (3) could be re-written as

$$\arg\max_{0\leq P_c<P}\left\{\sum_{\substack{i\in\Gamma_\omega^c\\i\neq c}}\eta_i\log_2\left(1+\frac{1}{\frac{P_cp_{rec,i}}{P^{RRM}P}+\frac{1}{SINR_i}}\right)+\eta_c\log_2\left(1+\frac{P_ch_c}{\sigma_c^2}\right)\right\} \quad (4)$$

Thus the optimization problem could be rewritten as

$$\arg\max_{0\leq P_c<P}\left\{\prod_{\substack{i\in\Gamma_\omega^c\\i\neq c}}\left(1+\frac{1}{\frac{P_cp_{rec,i}}{P^{RRM}P}+\frac{1}{SINR_i}}\right)^{\eta_i}+\left(1+\frac{P_ch_c}{\sigma_c^2}\right)^{\eta_c}\right\} \quad (5)$$

This is a GP model which could be solved with optimization method.

**2.4 resource usage extension**

When the fMS$_c$ receives wrong data from fBS$_c$, it will send RRM again. At this time, the target SINR in RRM would be set as zero. When fBS receives such kind of RRM, it will not do power control again and keeps the power control result as the last time.

**3. Conclusions**

In this paper an over the air resource coordination scheme is proposed for the femtocell networks. The proposed scheme reduces the network interferences while maintaining good femtocell indoor coverage. The next step of our work is to do the system level simulations to verify that the capacity of the system with the proposed resource coordination scheme could be improved than that of the conventional macrocell and femtocell network.